\documentclass[twocolumn,prb,aps,floatfix,showpacs]{revtex4}
\usepackage{graphicx}

\begin{document}
\draft
\title{Superconductivity from doping boron icosahedra.}
\author{Matteo Calandra}
\affiliation{Laboratoire de Min\'eralogie-Cristallographie, case 115, 4 place Jussieu, 75252, Paris cedex 05, France}
\author{Nathalie Vast}
\affiliation{Laboratoire des Solides Irradi\'es, CEA-CNRS-Ecole Polytechnique, 91128 Palaiseau, France}
\author{Francesco Mauri}
\affiliation{Laboratoire de Min\'eralogie-Cristallographie, case 115, 4 Place Jussieu, 75252, Paris cedex 05, France}
\date{\today}
\begin{abstract}
We propose a new route to achieve the superconducting state  in boron-rich solids, 
the hole doping of B$_{12}$ icosahedra.
For this purpose we consider a prototype metallic phase 
of B$_{13}$C$_2$. We show that in this compound the boron icosahedral units are 
mainly responsible for the large phonon frequencies logarithmic average, 
$\langle 65.8\rangle$ meV, and the moderate electron-phonon coupling $\lambda=0.81$. 
We suggest that this high $T_c$ could be a general feature of
hole doped boron icosahedral solids.
Moreover our calculated moderate value of $\lambda$ excludes the formation
of bipolarons localized on the icosahedral length scale as suggested by previous
authors.
\end{abstract}
\maketitle

\section{Introduction}

Low atomic number elements have been intensively investigated 
in the attempt of finding electron phonon mediated superconductors 
with high critical temperatures ($T_c$).  
Boron rich solids are eminent examples. 
The high energy phonon frequencies of metallic boron layers in the 
structure of magnesium diboride \cite{Nagamatsu} 
are mainly responsible for the 39 K $T_c$. 
Intercalation of these boron layers with lithium and 
boron substitution with carbon are currently under study 
\cite{Rosner} in the quest of even higher $T_c$.
Elemental boron becomes metallic (in a non-icosahedral
structure \cite{Eremets,loubeyre}) and superconducting at 
160 Gpa, with a $T_c$  which increases up to 11.2 K
at 250 Gpa \cite{Eremets}. 
A 23 K $T_c$ has recently been 
discovered in intermetallic, yttrium palladium boron carbides\cite{Cava}.

High temperature superconductivity has been found in alkali
intercalated fullerene (C$_{60}$) and theoretically suggested in other 
intercalated carbon polyhedrons,   
C$_{20}$\cite{Spagnolatti},  C$_{28}$\cite{Breda} or  C$_{36}$\cite{Collins}.
C$_{60}$ is a band insulator under normal conditions
but the intercalation with alkali atoms does generate a metallic state and a 
consequent superconducting state with $T_c$ up to $40 K$\cite{Gunnarssonrmp}.
Superconductivity is mainly sustained by the high frequency intramolecular
phonons so that most of the physical properties of 
alkali doped fullerenes can be understood from the solid C$_{60}$ electronic 
and phonon structures. 
In particular it is seen that the small radius of the molecule 
substantially increases the electron-phonon coupling with respect to
the case of the unrolled graphite layer \cite{Schluter}. 
The study of solids made of light atoms  
is then interesting since they can satisfy the
two important requirements of having large phonon frequencies and 
fairly high electron-phonon coupling. 

In this work we propose a new route to achieve the superconducting
state  in boron-rich solids, the hole doping of B$_{12}$ icosahedra.
For this purpose we consider a prototype metallic phase
of B$_{13}$C$_2$ (sec. \ref{sec:crystal}) for which we predict a $T_c$
comparable to that of MgB$_2$. We show (sec. \ref{sec:calcul}) that
in this compound the boron icosahedral units are
the main responsible for the predicted large $T_c$.
Thus this high $T_c$ is a general feature of
hole doped boron icosahedral solids. 

\begin{figure}[t]
\resizebox{3.0in}{2.5in}{\includegraphics{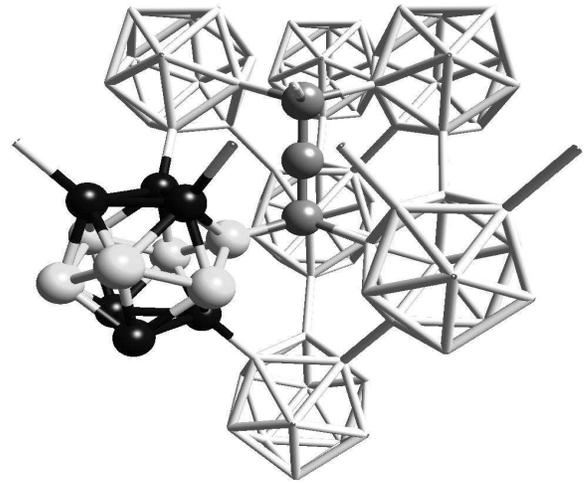}}
\caption{Atomic structure of B$_{12}$C$_3$ and 
of B$_{13}$C$_{2}$.  The black atoms are the
polar sites, bonded to neighboring icosahedra. The white atoms are the equatorial 
sites. The gray atoms form the chain. In B$_{12}C_3$ the atoms in the chain 
are CBC and the
icosahedra are B$_{11}$C with the carbon placed in a polar site. In B$_{13}$C$_{2}$
the carbon atom in the polar site is replaced with a boron atom.}
\label{fig:B4C}
\end{figure}

\section{\label{sec:crystal}Cristal structure}

We consider the hole doping of boron carbide B$_{12}$C$_3$ (or B$_{4}$C), a 
wide gap band-insulator.
Its crystal structure consists of an arrangement of
B$_{11}$C distorted icosahedra on the site of a rhombohedral lattice and of a linear
C-B-C atom chain \cite{Kirfel,Xray,Tallant,Kirkpatrick,Kuhlmann,Lazzari,Mauri}. The periodic unit cell contains 15 atoms and is illustrated in fig. \ref{fig:B4C}. 
Hole doped boron icosahedra can be naturally obtained replacing in B$_{12}$C$_{3}$
a carbon atom by a boron, giving B$_{13}$C$_{2}$. Due to the carbon 
substitution there is one hole per unit cell and,  B$_{13}$C$_{2}$ being non
magnetic, band theory predicts a metallic behaviour.  
At zero temperature and ambient pressure however, 
B$_{13}$C$_{2}$ is a semiconductor \cite{zuppiroli}.
The insulating character might be related to {\it (i)} the presence
of structural defects \cite{defects,thevenot} {\it (ii)} 
Mott polaronic features\cite{Wood}.
In the first case a metallic state might be achieved simply by producing high quality 
samples. In the second case it might be generated by the application 
of hydrostatic pressure in order to increase the hopping between the icosahedral
units.  The pressure necessary to metallise B$_{13}$C$_{2}$ would be 
lower than that necessary to metallise B$_{12}$C$_{3}$ or $\alpha-$B.
Indeed, the Mott-polaron gap in B$_{13}$C$_{2}$ is expected to be much
smaller than the several eV gaps of the band insulators B$_{12}$C$_{3}$ 
and $\alpha-$B.

The two most probable structures of B$_{13}$C$_{2}$ are 
B$_{11}$C(BBC), {\it i.e.} B$_{11}$C icosahedra 
linked by BBC chains, or B$_{12}$(CBC), {\it i.e.}  boron icosahedra linked 
by CBC chains.
From the theoretical point of view, Density Functional Theory (DFT) calculations 
\cite{Bylander} identified the B$_{12}$(CBC) structure 
as the most stable one with a 2.09 eV/cell larger binding energy 
with respect to B$_{11}$C(BBC).
Experimentally \cite{Kirkpatrick}
 the $^{13}$C  NMR spectrum of B$_{13}$C$_2$  confirms the DFT
calculations. Indeed changing composition from B$_{12}$C$_{3}$ to 
B$_{13}$C$_{2}$ the NMR peak associated to the C atom in the icosahedron
disappears. 
Therefore, in this paper,
we use the B$_{12}$(CBC) structure 
 and we refer to it as the B$_{13}$C$_{2}$ crystal structure. In other works
it has been suggested that the substituted boron is on the chain\cite{Aselageb}.
We did not consider this proposed structure since we do not 
expect that the conclusions presented in this work are substantially 
affected by the location of the additional B atom. 
\begin{figure}[t]
\resizebox{3.0in}{2.3in}{\includegraphics{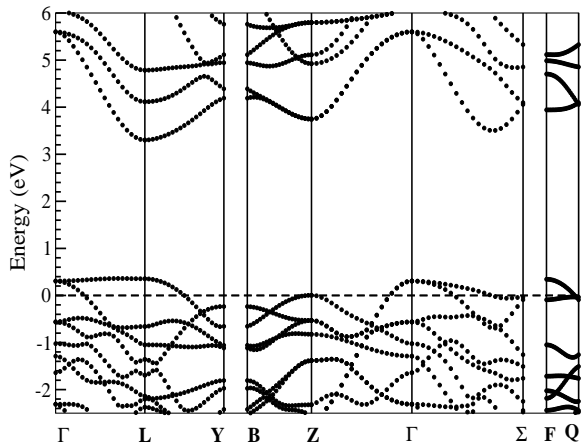}}
\caption{DFT band structure of B$_{13}$C$_2$. 
The energy (eV)  is referred to the Fermi energy level. The convention
for high symmetry points in the Brillouin zone is from ref. \onlinecite{Bradley}.}
\label{fig:bands}
\end{figure}

\section{\label{sec:calcul}Calculations and results}

In B$_{13}$C$_{2}$ a metallic phase could be achieved either by obtaining
clean samples or by applying a small pressure.
In DFT, already at room pressure, the ground state
is metallic. Therefore we use DFT to study the superconducting properties
of the hypothetical metallic state of B$_{13}$C$_{2}$.
Electronic structure calculations \cite{PWSCF} and geometrical optimization 
 are performed using DFT in the local density approximation. 
We use norm conserving pseudopotentials \cite{Troullier}
with a $s$  non local part. 
The wave-functions are expanded in plane waves using a $40$ Ry cutoff.
For the electronic structure  we sample 
the Brillouin zone (BZ) using a $4^3$ Monkhorst-Pack grid 
(10 ${\bf k}$-points in the irreducible BZ wedge)
 and  first order Hermite-Gaussian smearing 
of $0.03$ Ry. 
From geometrical optimization we obtain the values of
$a=9.686$ (a.u.) and  $\alpha=66.05$ (deg.) for the cell parameters of the
rhombohedral unit cell, very close to the experimental values \cite{Kirfel}
($a=9.823$ a.u. and $\alpha=65.62$ deg. ).

\begin{figure}[t]
\resizebox{3.0in}{2.3in}{\includegraphics{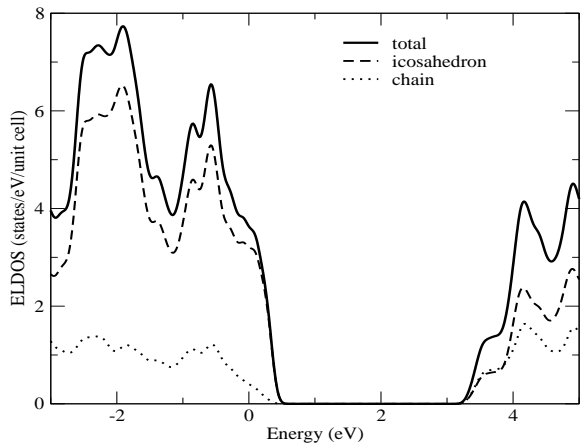}}
\caption{DFT electronic density of states of B$_{13}$C$_2$  
compared to the icosahedral atoms and chain atoms projected density of states.
The energy (eV)  is referred to the Fermi energy level.
}
\label{fig:eldos}
\end{figure}

The DFT band structure of B$_{13}$C$_2$  is shown in fig. \ref{fig:bands} and
is in good agreement with ref. \onlinecite{Bylander}.
It displays a metallic behaviour. The Fermi level is close to the top of the
valence band and it is crossed by several bands. 
We calculated the  electronic density of states (see fig. \ref{fig:eldos})
of B$_{13}$C$_{2}$, using a mesh of $N_k=14^3$ inequivalent 
${\bf k}-$points. The mesh is generated by taking the mesh
centered at the $\Gamma$ point and shifting it by a random
vector.
The density of states at the Fermi level is $N(0)=3.6$ states/eV/unit cell. 
We decompose the $N(0)$ in icosahedral and chain density of states,
by projecting the {\it ab initio} B$_{13}$C$_{2}$ 
wavefunction on the basis formed by the respective 
atomic pseudo-wavefunction (Lowdin population)\cite{sanchez}. 
At the Fermi level the icosahedral states are responsible
for 88\% of the total density of states ($N_{ico}(0)=3.2$ ). 
In particular the boron atoms in the polar sites have the largest contribution 
to $N_{ico}(0)$, namely $N_{polar}(0)=2.3$ and the
contribution due to the equatorial sites is smaller, $N_{eq}(0)=0.9$.
Thus, most of the electrons involved in conduction processes
resides in icosahedral states and only a small part in chain states,
($N_{chain}(0)=0.4$).

\begin{figure}
\includegraphics[height=2.5in,width=3.5in]{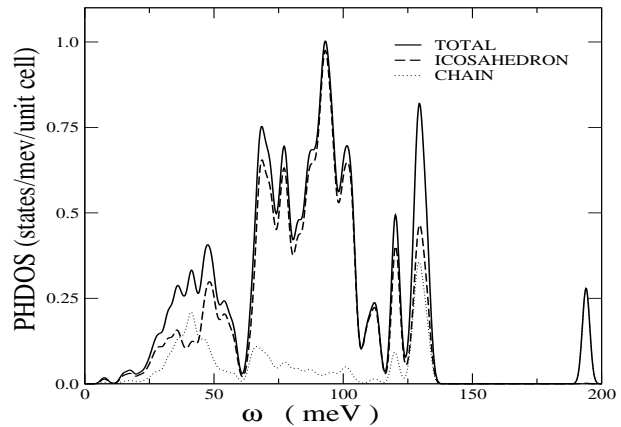}
\caption{Phonon density of states, $F(\omega)$, of B$_{13}$C$_2$.
The total density of states (solid line) compared to the phonon density 
of states of the icosahedral modes (dashed line) and of the 
chain modes (dotted line). Most of the weight
between 0 and 140 meV is due to icosahedral modes. }
\label{fig:phdos}
\end{figure}

We compute the harmonic phonon frequencies $\omega_{{\bf q},\nu}$ using
Density Functional Perturbation Theory in the linear response\cite{PWSCF}.
We use a  $N=4^3$  Monkhorst-Pack ${\bf q}-$points mesh, ${\bf q}$
being the phonon wavevector. The total phonon density of states, $F(\omega)$,
together with the phonon density of states restricted to the icosahedral 
and chain phonon modes are shown in fig. \ref{fig:phdos}.
The large number of peaks in $F(\omega)$ is determined by the large number
of phonon modes present in the system.

The icosahedral phonon modes are responsible for most of the 
 weight between 0 and 140 meV. The vibrations of the atoms in the
chain explain the high energy feature 
at 193 meV (both B and C vibrations) and part
of the feature at 129 meV (C vibrations). The structures in the 
chain restricted $F(\omega)$ between 25 and 55 meV are mainly due to B vibrations,
while the remaining weight between 60 and 106 meV 
is due to C vibrations.

The electron-phonon interaction for a phonon mode $\nu$ with momentum $q$ 
can be written as:
\begin{equation}\label{eq:elph}
\lambda_{{\bf q}\nu} = \frac{4}{\omega_{{\bf q}}N(0) N_{k}} \sum_{{\bf k},n,m} |g_{{\bf k}n,{\bf k+q}m}^{\nu}|^2 \delta(\varepsilon_{{\bf k}n}) \delta(\varepsilon_{{\bf k+q}m})
\label{eq:lambdaq}
\end{equation}
where the sum is carried out over the BZ, 
and $\varepsilon_{{\bf k}n}$ are the energy bands measured with
respect to the Fermi level at point ${\bf k}$.
The matrix element is
$g_{{\bf k}n,{\bf k+q}m}^{\nu}= \langle {\bf k}n|\delta V/\delta u_{{\bf q}\nu} |{\bf k+q} m\rangle /\sqrt{2 \omega_{{\bf q}\nu}}$,
where $u_{{\bf q}\nu}$ is the amplitude of the displacement of the phonon $\nu$
of wavevector ${\bf q}$, $V$ is the Kohn-Sham
potential and $N(0)=3.6$ states/eV/unit cell is the electronic DOS at the Fermi level.
The electron-phonon coupling $\lambda$ is then
calculated as an average over the $N$ ${\bf q}-$points
mesh and over all the modes,  
$\lambda=\sum_{{\bf q}\nu} \lambda_{{\bf q}\nu}/N = 0.81$.

\begin{figure}[ht]
\includegraphics[height=2.5in,width=3.5in]{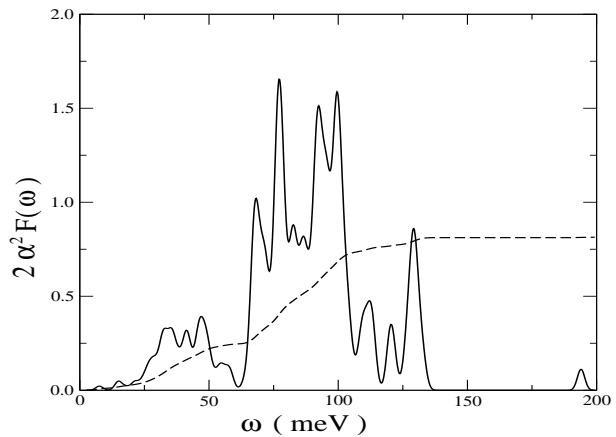}
\caption{Eliashberg
function (solid line) and average electron-phonon coupling (dashed line) 
of B$_{13}$C$_2$.The icosahedral phonon modes ranging between 60 and 105 meV
give the largest contribution to $\lambda$.}
\label{fig:eliash}
\end{figure}

The modes responsible for superconductivity can be identified
from the Eliashberg function 
$\alpha^2F(\omega)$
\begin{equation}
\alpha^2F(\omega)=\frac{1}{2 N}\sum_{{\bf q}\nu} \lambda_{{\bf q}\nu} \omega_{{\bf q}\nu} \delta(\omega-\omega_{{\bf q}\nu} )
\end{equation}
The Eliashberg function is depicted in fig. \ref{fig:eliash}. 
Most of the contribution to $\lambda$ comes from the region 
from 60 to 105 meV, mainly
related to the icosahedral phonon modes. 

The critical superconducting temperature is estimated from the calculated
phonon frequencies and electron-phonon coupling using the McMillan formula\cite{mcmillan}:
\begin{equation}
T_c = \frac{\langle \omega \rangle}{1.2} \exp\left( - \frac{1.04 (1+\lambda)}{\lambda-\mu^* (1+0.62\lambda)}\right)\label{eq:mcmillan}
\end{equation}
where $\mu^*$ is the screened Coulomb pseudopotential which takes 
into account the Coulomb repulsion between the electrons dressed by retardation effects
due to the phonons and $\langle\omega\rangle=65.8$ meV is the phonon frequencies 
logarithmic average. 
The calculated values of $T_c$ for B$_{13}$C$_2$ as a function of $\mu^{*}$ are 
illustrated in  table \ref{tabella3}. 
The critical temperature for metallic
B$_{13}$C$_2$ is comparable to the one obtained for MgB$_{2}$ and, using the McMillan
formula, ranges between  15.8 k and 36.7 K.

\begin{table}[hbt]
\begin{ruledtabular}
\begin{tabular}{lcccc} 
Material & $\mu^{*}$ & $\langle \omega \rangle $  (meV)  & $\lambda$ & $T_c$ (K) \\ \hline 
B$_{13}$C$_{2}$ &  0.1  & 65.8 & 0.81 & 36.7 \\  
                & 0.14 & 65.8 & 0.81  & 27.6 \\  
                & 0.2  & 65.8 & 0.81  & 15.8 \\  
MgB$_{2}$ & 0.14 & 62.0 & 0.87 & 30.7 
\end{tabular}
\end{ruledtabular}
\caption{Critical temperatures of B$_{13}$C$_{2}$ and MgB$_2$. Predicted critical 
temperatures as a function of the screened Coulomb pseudopotential ($\mu^{*}$). 
The critical temperature is estimated
using the McMillan formula (eq. \ref{eq:mcmillan}).The results are compared with MgB$_{2}$ 
(from ref. \protect\cite{Kong}).}
\label{tabella3}
\end{table}

Our calculation of the electron-phonon coupling gives also new insight on the 
possible occurrence of a bipolaronic insulating state in B$_{13}$C$_{2}$,
as suggested by other authors \cite{Wood}.
In icosahedral boron compounds the bipolaron could be localized 
on two length scales: the icosahedron length scale or a smaller one, 
of the order of the bond length. 
If the bipolaron were localized on the icosahedron 
length scale (as most of the bipolaronic literature seems to suggest
\cite{Wood})
the bipolaronic distortion could be seen as a small perturbation
to the metallic state. 
In this case the bipolaron could be described within a model considering
a linear electron-phonon coupling perturbation
to the harmonic Hamiltonian in the metallic phase.
Such a model is the one used in the present paper
to study the superconducting properties.
Our calculated value of $\lambda$ is too small\cite{Robasc} to justify
a bipolaronic insulating state for B$_{13}$C$_{2}$.
On the contrary our calculation cannot exclude the occurrence
of a bipolaron localized  on the bond-length scale.
In this scenario the bipolaron would involve a  
substantial deformation of our crystal structure (e.g. the
breaking or formation of chemical bonds). Such a strong 
deformation could not be seen as a weak perturbation 
to the harmonic Hamiltonian in the metallic state and
could not be described by the approach used in the present paper.

\section{Conclusions}

In this work we have studied the possible occurrence of superconductivity from hole
doping boron icosahedra.
We found the possibility of having high superconducting critical
temperatures in these systems. 
As a possible physical realization of
a metallic state we have considered  B$_{13}$C$_{2}$, which is formed from
B$_{12}$ icosahedral units with one hole per icosahedra. Using {\it ab-initio}
calculations we have determined its normal state properties, 
finding a moderate electron-phonon coupling and a large phonon
logarithmic average to the phonon frequencies.
We have demonstrated that both properties are connected
to the B$_{12}$  building blocks. Indeed 
the local density of state at the Fermi level 
and the  phonon modes strongly coupled with electrons
are localized on the icosahedra.
As a consequence our findings are not restricted to  B$_{13}$C$_{2}$ compounds but
can be applied to other metallic compounds composed
by B rich icosahedra. For example another possibility to achieve a 
metallic state 
is the substitution of P with Si
in B$_{12}$P$_{2}$   or of As with Si in B$_{12}$As$_{2}$. 
B$_{12}$P$_{2}$ and B$_{12}$As$_{2}$ are band insulators composed by
B$_{12}$ icosahedra and 2-atom P$_{2}$ or As$_2$ chains\cite{Aselageb}. 
The substitution of pentavalent atoms like As and P with a tetravalent
Si introduces a hole in the system. B$_{12}$P$_{2-x}$Si$_{x}$ wafer 
resistivity measurements show a low
electrical conductivity \cite{Kumashiro}. If such conduction were related to
bipolaron formation then a metallic state could be achieved applying 
pressure.  An advantage of the 2-atom chain systems 
(B$_{12}$P$_{2-x}$Si$_{x}$ or B$_{12}$As$_{2-x}$Si$_{x}$) respect to the
3-atom chain systems (B$_{13}$C$_{2}$) is the lack in the formers of 
an internal soft degree of freedom,
the chain bending. Probably this feature makes
the 2-atom chain structures more stable under pressure. 

\section{Acknowledgements}

We acknowledge illuminating discussion with M. Bernasconi, P. Giannozzi and E. Tosatti.
We thank C. J. Pickard for providing us the figure of the molecular structure.
Computer time has been granted by IDRIS (project 000544 and 021202).
M.C. was supported by a Marie Curie Fellowship of the European Commission, 
contract No. IHP-HPMF-CT-2001-01185.

\end{document}